\documentclass{article}
\usepackage[latin9]{inputenc}
\usepackage{color}
\usepackage{array}
\usepackage{verbatim}
\usepackage{units}
\usepackage{multirow}
\usepackage{amsmath}
\usepackage{graphicx}
\usepackage[unicode=true]
 {hyperref}
\usepackage{breakurl}

\makeatletter

\providecommand{\tabularnewline}{\\}

\usepackage{ismir}
\usepackage{cite}\usepackage{url}

\title{Metrical-accent aware vocal onset detection in polyphonic audio}

\multauthor
{  Georgi Dzhambazov$^1$\hspace{1.5cm} Andre Holzapfel$^2$} { \bfseries{ Ajay Srinivasamurthy$^1$ \hspace{1.8cm}Xavier Serra$^1$}\\
$^1$  Music Technology Group, Universitat Pompeu Fabra, Barcelona, Spain\\
$^2$ Media Technology and Interaction Design, KTH Royal Institute of Technology, Stockholm, Sweden\\
{\tt\small \{georgi.dzhambazov,ajays.murthy,xavier.serra\}@upf.edu, holzap@kth.se}\\
}
\def\authorname{Georgi Dzhambazov, Andre Holzapfel, Ajay Srinivasamurthy, Xavier Serra}

\@ifundefined{showcaptionsetup}{}{%
 \PassOptionsToPackage{caption=false}{subfig}}
\usepackage{subfig}
\makeatother

\begin{document}
\maketitle 
\begin{abstract}
The goal of this study is the automatic detection of onsets of the
singing voice in polyphonic audio recordings. Starting with a hypothesis
that the knowledge of the current position in a metrical cycle (i.e.
metrical accent) can improve the accuracy of vocal note onset detection,
we propose a novel probabilistic model to jointly track beats and
vocal note onsets. The proposed model extends a state of the art model
for beat and meter tracking, in which a-priori probability of a note
at a specific metrical accent interacts with the probability of observing
a vocal note onset. We carry out an evaluation on a varied collection
of multi-instrument datasets from two music traditions (English popular
music and Turkish makam) with different types of metrical cycles and
singing styles. Results confirm that the proposed model reasonably
improves vocal note onset detection accuracy compared to a baseline
model that does not take metrical position into account.
\end{abstract}

\section{Introduction}

Singing voice analysis is one of the most important topics in the
field of music information retrieval because singing voice often forms
the melody line and creates the impression of a musical piece. The
automatic transcription of singing voice can be considered to be a
key technology in computational studies of singing voice. It can be
utilized for end-user applications such as enriched music listening
 and singing education. It can as well enable other computational
tasks including singing voice separation, karaoke-like singing voice
suppression or lyrics-to-audio alignment \cite{goto2014singing}. 

The process of converting an audio recording into some form of musical
notation is commonly known as automatic music transcription. Current
transcription methods use general purpose models, which are unable
to capture the rich diversity found in music signals \cite{benetos2013automatic}.
In particular, singing voice poses a challenge to transcription algorithms
because of its soft onsets, and phenomena such as portamento and vibrato.
One of the core subtasks of singing voice transcription (SVT) is detecting
note events with a discrete pitch value, an onset time and an offset
time from the estimated time-pitch representation. Detecting the time
locations of vocal note onsets can benefit from automatically detected
events from musical facets, such as musical meter. \textcolor{black}{In
fact, the accents in the metrical cycle determine to a large extent
the temporal backbone of singing melody lines. }Studies on sheet music
showed that the locations of vocal note onsets are influenced by the
their position in a metrical cycle \cite{huron2006sweet,holzapfel2015relation}.
Despite that, there have been few studies on meter aware analysis
of onsets in music audio \cite{degara2010note}. 

In this work we propose a novel probabilistic model that tracks simultaneously
note onsets of singing voice and instrumental energy accents in a
metrical cycle. We extend a state of the art model for beat and meter
tracking, based on dynamic Bayesian networks (DBN). A model variable
is added that models the temporal segments of a note and their interaction
with metrical position. The proposed model is applied for the automatic
detection of vocal note onsets in multi-instrumental recordings with
predominant singing voice. Evaluation is carried out on datasets from
music traditions, for which there is a clear correlation between metrical
accents and the onset times in the vocal line.

\section{Related Work}

\subsection{Singing voice transcription}

 A probabilistic note hidden Markov model (HMM) is presented in \cite{ryynanen2004probabilistic},
where a note has 3 states: attack (onset), stable pitch state and
silent state. The transition probabilities are learned from data.
 Recently \cite{mauch2015computer} suggested to compact musical
knowledge into rules as a way to describe the observation and transition
likelihoods, instead of learning them from data. The authors suggest
covering a range with distinct pitch from lowest MIDI C2 up to B7.
\textcolor{black}{Each MIDI pitch is further divided into 3 sub-pitches,
resulting in $n=207$ notes with different pitch, each having the
3 note states. }Although being conceptually capable of tracking onsets
in singing voice audio with accompaniment, these approaches were tested
only on a cappella singing. 

In multi-instrumental recordings, an essential first step is to extract
reliably the predominant vocal melody. There have been few works dealing
with SVT in multi-instrumental recordings in general \cite{kroher2015automatic,NishikimiNIY16},
and with onset detection, in particular \cite{chang2014pairwise}.
Some of them \cite{kroher2015automatic,NishikimiNIY16} rely on the
algorithm  for predominant melody extraction of \cite{salamon2012melody}.
\textcolor{red}{}

\subsection{Beat Detection}

Recently a Bayesian approach, referred to as the \emph{bar-pointer}
model, has been presented \cite{whiteley:06:ismir}. It describes
events in music as being driven by their current position in a metrical
cycle (i.e. musical bar). \textcolor{black}{The model represents as
hidden variables in a }Dynamic Bayesian network (DBN)\textcolor{black}{{}
the current position in a bar, the tempo, and the type of musical
meter, which can be referred to as bar-tempo state space.}

The work of \cite{holzapfel2014tracking} applied this model to recordings
from non-Western music, in order to handle jointly beat and downbeat
tracking. The authors showed that the original model can be adapted
to different rhythmic styles and time signatures, and an evaluation
is presented on Indian, Cretan and Turkish music datasets. 

\textcolor{black}{Later \cite{krebs:15:ismir} suggested a modification
of the bar-tempo state space, in order to reduce the computational
burden from its huge size. }\textcolor{red}{ }

\section{Datasets}

\subsection{Turkish makam\label{subsec:Turkish-makam}}

The Turkish dataset has two meter types, referred to as usuls in Turkish
makam: the 9/8-usul aksak and the 8/8-usul d{\"u}yek. It is a subset
of the dataset presented in \cite{holzapfel2014tracking}, including
only the recordings with singing voice present. The beats and downbeats
were annotated by \cite{holzapfel2014tracking}. The vocal note onsets
are annotated by the first author, whereby only pitched onsets are
considered (2100 onsets). To this end, if a syllable starts with an
unvoiced consonant, the onset is placed at the beginning of the succeeding
voiced phoneme\footnote{The dataset is described at \href{http://compmusic.upf.edu/node/345}{http://compmusic.upf.edu/node/345}}. 

For this study we divided the dataset into training and test subsets.
The test dataset comprises 5 1-minute excerpts from recordings with
solo singing voice only for each of the two usuls (on total 780 onsets).
The training dataset spans around 7 minutes of audio from each of
the two usuls. Due to the scarcity of material with solo singing voice,
several excerpts with choir sections were included in the training
data.

\subsection{English pop}

The datasets, on which singing voice transcription in multi-instrumental
music is evaluated, are very few \cite{benetos2013automatic}: Often
a subset of the RWC dataset is employed, which does not contain diverse
genres and singers \cite{Goto02rwcmusic}. To overcome this bias,
we compiled the \emph{lakh-vocal-segments} dataset: We selected 14
30-second audio clips of English pop songs, which have been aligned
to their corresponding MIDIs in a recent study \cite{raffel2016learning}.
Criteria for selecting the clips are the predominance of the vocal
line; 4/4 meter; correlation between the beats and the onset times.
We derived the locations of the vocal onsets (850 on total) from the
aligned vocal MIDI channel, whereby some imprecise locations were
manually corrected. To encourage further studies on singing voice
transcription we make available the derived annotations\footnote{\href{https://github.com/georgid/lakh_vocal_segments_dataset}{https://github.com/georgid/lakh\_{}vocal\_{}segments\_{}dataset}}. 

\section{Approach}

\begin{figure}
\begin{centering}
\includegraphics[width=1\columnwidth]{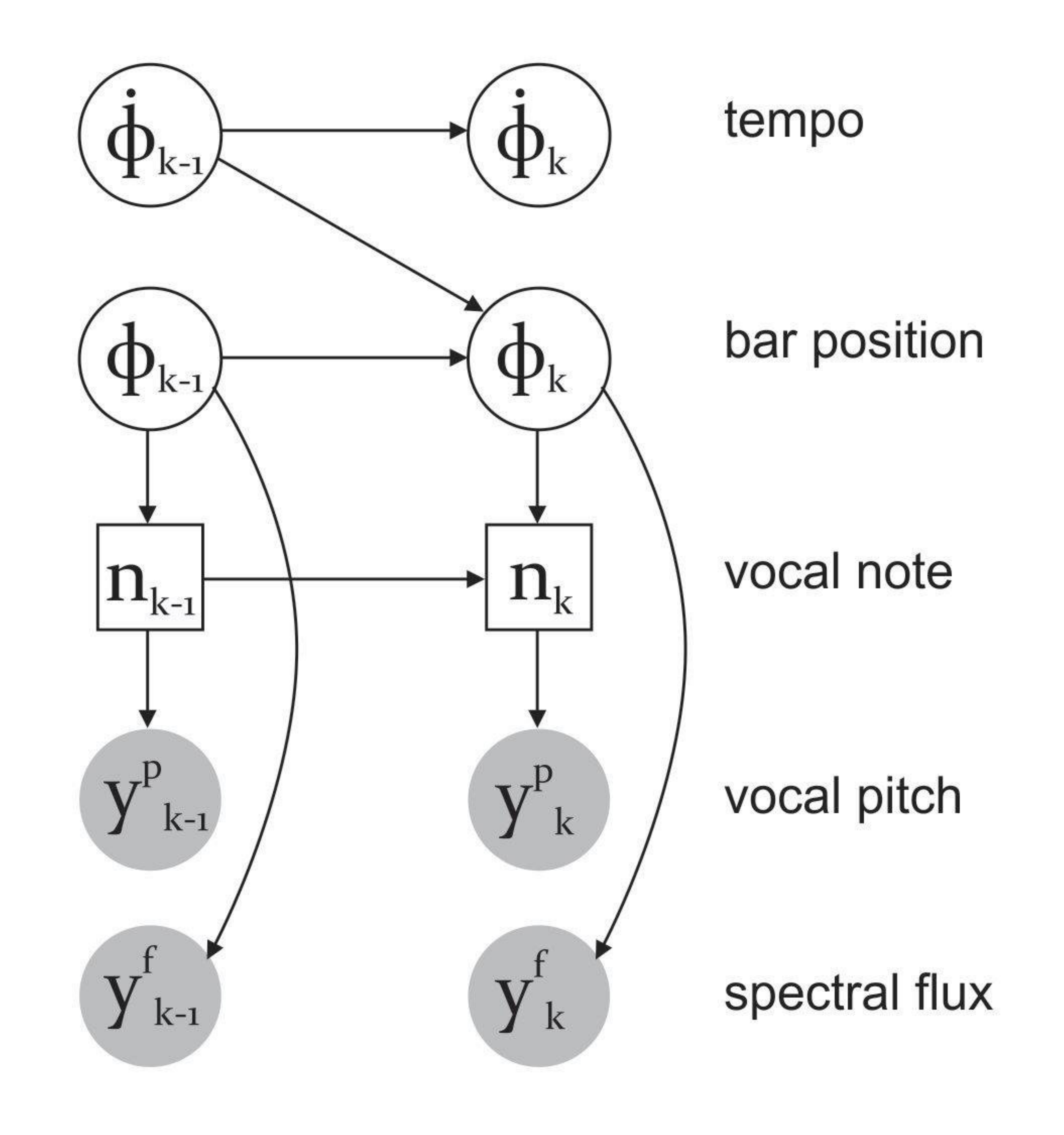}
\par\end{centering}
\caption{A dynamic Bayesian network for the proposed beat and vocal onset detection
model. Circles and squares denote continuous and discrete variables,
respectively. Gray nodes and white nodes represent observed and hidden
variables, respectively.}
\label{dbn}
\end{figure}

The proposed approach extends the beat and meter tracking model, presented
in \cite{krebs:15:ismir}. \textcolor{black}{We adopt from it the
variables for the position in the metircal cycle (bar position) $\phi$
and the instantaneous tempo }$\dot{\phi}$. We also adopt the observation
model, which describes how\textcolor{black}{{} the metrical accents
(beats) are related to an observed onset feature vector $y_{f}$.
}All variables and their conditional dependencies are represented
as the hidden variables in a DBN (see Figure \ref{dbn}). We consider
that the \emph{a priori} probability of a note at a specific metrical
accent interacts with the probability of observing a vocal note onset.
To represent that interaction we add a hidden state for the temporal
segment of a vocal note \emph{n}, which depends on the current position
in the metrical cycle. The probability of observing a vocal onset\textcolor{black}{{}
is derived from the emitted pitch $y_{p}$ of the vocal melody.} 

In the proposed DBN, an observed sequence of features derived from
an audio signal $y_{1:K}=\{y,..,y_{K}\}$ is generated by a sequence
of hidden (unknown) variables $x_{1:K}=\{x_{1},...,x_{K}\}$, where
K is the length of the sequence (number of audio frames in an audio
excerpt). The joint probability distribution of hidden and observed
variables factorizes as: 
\begin{equation}
P(x_{1:K},y_{1:K})=P(x_{0})\Pi_{k=1}^{K}P(x_{k}|x_{k-1})P(y_{k}|x_{k})\label{eq:inference}
\end{equation}

where $P(x_{0})$ is the initial state distribution; $P(x_{k}|x_{k-1})$
is the transition model and $P(y_{k}|x_{k})$ is the observation model.

\subsection{Hidden variables}

At each audio frame $k$, the hidden variables describe the state
of a hypothetical bar pointer $x_{k}=[\dot{\phi_{k}},\phi_{k},n_{k}]$,
representing the instantaneous tempo, the bar position and the vocal
note respectively. 

\subsubsection{Tempo state $\dot{\phi}$ and bar position state $\phi$ }

\textcolor{black}{The bar position $\phi$ points to the current position
in the metrical cycle (bar). The instantaneous tempo $\dot{\phi}$
encodes how many bar positions the pointer advances from the current
to the next time instant. To assure feasible computational time we
relied on the combined bar-tempo efficient state space, presented
in \cite{krebs:15:ismir}. To keep the size of the bar-tempo state
space small, we input the ground truth tempo for each recording, allowing
a range for }$\dot{\phi}$\textcolor{black}{{} within $\pm10$ bpm from
it, in order to accommodate gradual tempo changes. This was the minimal
margin at which beat tracking accuracy did not degrade substantially.
For a study with data with higher stylistic diversity, it would make
sense to increase it to at least 20\% as it is done in \cite[Section 5.2]{holzapfel2016bayesian}.
This yields around 100-1000 states for the bar positions within a
single beat (in the order of $5000$ for 4 beats, and $10000$ for
8-9 beats for the usuls ).}

\subsubsection{Vocal note state $n$\label{subsec:Note-state}}

The vocal note states represent the temporal segments of a sung note.
They are a modified version of these suggested in the note transcription
model of \cite{mauch2015computer}. We adopted the first two segments:
attack region (A), stable pitch region (S). We replaced the silent
segment with non-vocal state (N). Because full-fledged note transcription
is outside the scope of this work, instead of 3 steps per semitone,
we used for simplicity only a single one, which deteriorated just
slightly the note onset detection accuracy. Also, to reflect the pitch
range in the datasets, on which we evaluate, we set as minimal MIDI
note E3 covering almost 3 octaves up to B5 (35 semitones). This totals
to 105 note states. 

\medskip{}

To be able to represent the DBN as an HMM, the bar-tempo efficient
state space is combined with the note state space into a joint state
space \emph{x}. The joint state space is a cartesian product of the
two state spaces, resulting in up to $10000$$\times105\approx1\thinspace$M
states.

\subsection{Transition model}

Due to the conditional dependence relations in Figure \ref{dbn} the
transitional model factorizes as

\begin{equation}
\begin{array}{cc}
P(x_{k}|x_{k-1})= & P(\dot{\phi}_{k}|\dot{\phi}_{k-1})\thinspace\times\\
P(\phi_{k}|\phi{}_{k-1},\dot{\phi}_{k-1})\thinspace\times & P(n_{k}|n_{k-1},\phi_{k})
\end{array}
\end{equation}

The tempo transition probability $p(\dot{\phi_{k}}|\dot{\phi}_{k-1})$
and bar position probability $p(\phi_{k}|\phi_{k-1},\dot{\phi}_{k-1})$
are the same as in\textcolor{black}{{} \cite{krebs:15:ismir}}\textcolor{blue}{.}\textcolor{black}{{}
Transition from one tempo to another is allowed only at bar positions,
at which the beat changes. This is a reasonable assumption for the
local tempo deviations in the analyzed datasets, which can be considered
to occur relatively beat-wise. }

\subsubsection{Note transition probability \label{par:Note-transition-probability}}

The probability of advancing to a next note state is based on the
transitions of the note-HMM, introduced in \cite{mauch2015computer}.
Let us briefly review it: From a given note segment the only possibility
is to progress to its following note segment. To ensure continuity
each of the self-transition probabilities is rather high, given by
constants \emph{$c_{A}$, $c_{S}$ }and \emph{$c_{N}$ }for A, S and
N segments respectively ($c_{A}$=0.9; $c_{S}$=0.99; $c_{N}=0.9999$).
Let $P_{N_{i}A_{j}}$ be the probability of transition from non-vocal
state $N_{i}$ after note $i$ to attack state $A_{j}$ of its following
note $j$.\textcolor{red}{{} }The authors assume that it depends on
the difference between the pitch values of notes $i$ and $j$ and
it can be approximated by a normal distribution centered at change
of zero (\cite{mauch2015computer}, Figure 1.b). This implies that
small pitch changes are more likely than larger ones. Now we can formalize
their note transition as:

\begin{equation}
p(n_{k}|n_{k-1})=\begin{cases}
P_{N_{i}A_{j}}, & n_{k-1}=N_{i}\quad n_{k}=A_{j}\\
c_{N}, & n_{k-1}=n_{k}=N_{i}\\
1-c_{A}, & n_{k-1=}A_{i}\quad n_{k}=S_{j}\\
c_{A}, & n_{k-1}=n_{k}=A_{i}\\
1-c_{S} & n_{k=1}=S_{i\quad}n_{k}=N_{j}\\
c_{S}, & n_{k-1}=n_{k}=S_{i}\\
0 & else
\end{cases}\label{eq:note_trans_probs}
\end{equation}

Note that the outbound transitions from all non-vocal states $N_{i}$
should sum to 1, meaning that
\begin{equation}
c_{N}=1-\sum_{i}P_{N_{i}A_{j}}\label{eq:normalization_trans_prob}
\end{equation}

\textcolor{black}{In this study, we modify $P_{N_{i}A_{j}}$ to allow
variation} in time, depending on the current bar position $\phi_{k}$. 

\begin{equation}
p(n_{k}|n_{k-1,}\phi_{k})=\begin{cases}
P_{N_{i}A_{j}}\Theta(\phi_{k}), & n_{k-1}=N_{i},n_{k}=A_{j}\\
c_{N}, & n_{k-1}=n_{k}=N_{i}\\
...
\end{cases}\label{eq:note_onset_trans_prob}
\end{equation}

where
\begin{description}
\item [{$\Theta(\phi_{k}):$}] function weighting the contribution of a
beat adjacent to current bar position $\phi_{k}$
\end{description}
and

\begin{equation}
c_{N}=1-\Theta(\phi_{k})\sum_{i}P_{N_{i}A_{j}}
\end{equation}

\textcolor{black}{The transition probabilities in all the rest of
the }cases remain the same. We explore two variants of the weighting
function $\Theta(\phi_{k}):$ 

\textbf{1. Time-window redistribution weighting:} Singers often advance
or delay slightly note onsets off the location of a beat. The work
\cite{NishikimiNIY16} presented an idea on how to model vocal onsets,
time-shifted from a beat, by stochastic distribution. Similarly, we
introduce a normal distribution $\mathcal{N}{}_{0,\sigma}$, centered
around 0 to re-distribute the importance of a metrical accent (beat)
over a time window around it. Let $b_{k}$ be the beat, closest in
time to a current bar position $\phi_{k}$. Now:

\begin{equation}
\Theta(\phi_{k})=[\mathcal{N}_{0,\sigma}(d(\phi_{k},b_{k}))]^{w}e(b_{k})\label{eq:weight_anno}
\end{equation}

where
\begin{description}
\item [{$e(b):$}] probability of a note onset co-occurring with the $b^{th}$
beat (b $\in$$B$); $B$ is the number of beats in a metrical cycle
\item [{$w:$}] sensitivity of vocal onset probability to beats 
\item [{$d(\phi_{k},b_{k}):$}] the distance from current bar position
$\phi_{k}$ to the position of the closest beat $b_{k}$ 
\end{description}
Equation \ref{eq:note_onset_trans_prob} means essentially that the
original $P_{N_{i}A_{j}}$ is scaled according to how close in time
to a beat it is. %

\textbf{2. Simple weighting:} We also aim at testing a more conservative
hypothesis that it is sufficient to approximate the influence of metrical
accents only at the locations of beats. To reflect that, we modify
the $P_{N_{i}A_{j}}$ only at bar positions corresponding to beat
positions, for which the weighting function is set to the peak of
$N_{0,\sigma}$, and to 1 elsewhere.

\begin{equation}
\Theta(\phi_{k})=\begin{cases}
[N_{0,\sigma}(0)]^{w}e(b_{k}), & d(\phi_{k},b_{k})=0\\
1 & else
\end{cases}\label{eq:weight_detected}
\end{equation}

\subsection{Observation models}

The observation probability $P(y_{k}|x_{k})$ describes the relation
between the hidden states and the (observed) audio signal. In this
work we make the assumption that the observed vocal pitch and the
observed metrical accent are conditionally independent from each other.
This assumption may not hold in cases when energy accents of singing
voice, which contribute to the total energy of the signal, are correlated
to changes in pitch. However, for music with percussive instruments
the importance of singing voice accents is diminished to a significant
extent by percussive accents. Now we can rewrite Eq. \ref{eq:inference}
as 

\begin{equation}
\begin{array}{cc}
P(x_{1:K},y_{1:K}^{f},y_{1:K}^{p})=\\[3pt]
P(x_{0})\Pi_{k=1}^{K}P(x_{k}|x_{k-1})P(y_{k}^{f}|x_{k})P(y_{k}^{p}|x_{k})
\end{array}
\end{equation}This means essentially that the observation probability can be represented
as the product of the observation probability of a metrical accent
$P(y_{k}^{f}|x_{k})$ and the observation probability of vocal pitch
\textcolor{black}{$P(y_{k}^{p}|x_{k})$}.

\subsubsection{Accent observation model}

In this paper for $P(y_{k}^{f}|x_{k})$ we train GMMs on the spectral
flux-like feature $y^{f}$, extracted from the audio signal using
the same parameters as in \cite{krebs:15:ismir} and \cite{holzapfel2014tracking}.
The feature vector $y^{f}$ summarizes the energy changes (accents)
that are likely to be related to the onsets of all instruments together.
This forms a rhythmic pattern of the accents, characteristic for a
given metrical type. The probability of observing an accent thus depends
on the position in the rhythmic pattern, $P(y_{k}^{f}|x_{k})=P(y_{k}^{f}|\phi_{k})$.

\subsubsection{Pitch observation model}

\textcolor{black}{The pitch probability $P(y_{k}^{p}|x_{k})$ }reduces
to \textcolor{black}{$P(y_{k}^{p}|n_{k})$}, because it depends only
on the current vocal note state. We adopt the idea proposed in \cite{mauch2015computer}
that a note state emits pitch $y^{p}$ according to a normal distribution,
centered around its average pitch. The standard deviation of stable
states and the one of the onset states are kept the same as in the
original model, respectively 0.9 and 5 semitones. The melody contour
of singing is extracted in a preprocessing step. We utilized for English
pop a method for predominant melody extraction \cite{salamon2012melody}.
For Turkish makam, we instead utilized an algorithm, extended from
\cite{salamon2012melody} and tailored to Turkish makam \cite{atli2014audio}.
In both algorithms, each audio frame $k$ gets assigned a pitch value
and probability of being voiced $v_{k}$ %
. Based on frames with zero probabilities, one can infer which segments
are vocal and which not. Since correct vocal segments is crucial for
the sake of this study and the voicing estimation of these melody
extraction algorithms are not state of the art, we manually annotated
segments with singing voice, and thus assigned $v_{k}=0$ for all
frames, annotated as non-vocal.

\textcolor{black}{For each state the observation probability $P(y_{k}^{p}|n_{k})$
of vocal states is normalized to sum to $v_{k}$ (unlike the original
model which sums to a global constant v).} This leaves the probability
for each non-vocal state be $\nicefrac{1-v_{k}}{n}$.

\subsection{Learning model parameters}

\subsubsection{Accent observation model\label{subsec:Bar-Observation-model}}

We trained the metrical accent probability $P(y_{k}^{f}|\phi_{k})$
separately for each meter type: The Turkish meters are trained on
the training subset of the makam dataset (see section \ref{subsec:Turkish-makam}).
For each usul (8/8 and 9/8) we trained a rhythmic pattern by fitting
a 2-mixture GMM on the extracted feature vector $y^{f}$. Analogously
to \cite{krebs:15:ismir}, we pooled the bar positions down to 16
patterns per beat. For English pop we used the 4/4 rhythmic pattern,
trained by \cite{krebs2013rhythmic} on ballroom dances. The feature
vector is normalized to zero mean, unit variance and taking moving
average. Normalization is done per song. 

\subsubsection{Probability of note onset}

The probability of a vocal note onset co-occurring at a given bar
position $e(b)$ is obtained from studies on sheet music.\textbf{
}Many notes are aligned with a beat in the music score, meaning a
higher probability of a note at beats compared to inter-beat bar positions.
A separate distribution $e(b)$ is applied for each different metrical
cycle. For the Turkish usuls $e(b)$ has been inferred from a recent
study \cite[Figure 5. a-c]{holzapfel2015relation}. The authors used
a corpus of music scores, on data from the same corpus, from which
we derived the Turkish dataset. The patterns reveal that notes are
expected to be located with much higher likelihoods on those beats
with percussive strokes than on the rest. 

In comparison to a classical tradition like makam, in modern pop music
the most likely positions of vocal accents in a bar are arguably much
more heterogeneous, due to the big diversity of time-deviations from
one singing style to another \cite{huron2006sweet}. Due to lack of
a distribution pattern $e(b)$, characteristic for English pop, we
set it manually with probabilities $(0.8,0.6,0.8,0.6)$ for the 4
beats.

\subsection{Inference}

We obtain the most optimal state sequence $x_{1:K}$ by decoding with
the Viterbi algorithm. A note onset is detected when the state path
enters an attack note state after being in non-vocal state.

\subsubsection{With manually annotated beats}

We explored the option that beats are given as input from a preprocessing
step (i.e. when they are manually annotated). In this case, the detection
of vocal onsets can be carried out by a reduced model with a single
hidden variable: the note state. The observation model is then reduced
to the pitch observation probability. The transition model is reduced
to a bar-position aware transition probability $a_{ij}(k)=p(n_{k}=j|n_{k-1}=i,\phi_{k})$
(see Eq. \ref{eq:note_onset_trans_prob}).\textcolor{black}{{} }To represent
the\textcolor{black}{{} time-dependent self-transition probabilities
we utilize time-varying transition matrix.  The standard transition
probabilities in the Viterbi maximization step are substituted for
the bar-position aware transitions $a_{ij}(k)$ }

\textcolor{black}{
\begin{equation}
\delta_{k}(j)=\max_{i\in(j,\thinspace j-1)}\delta_{k-1}(i)\thinspace a_{ij}(k)\thinspace b_{j}(O_{k})
\end{equation}
Here $b_{j}(O_{k})$ is the observation probability for state $i$
for feature vector $O_{k}$ and $\delta_{k}(j)$ is the probability
for the path with highest probability ending in state $j$ at time
$k$ (complying with the notation of \cite[III. B]{rabiner1989tutorial}}

\subsubsection{Full model}

In addition to onsets, a beat is detected when the bar position variable
hits one of $B$ positions of beats within the metrical cycle. 

Note that the size of the state space $x$ poses a memory requirement.
A recording of 1 minute has around $10000$ frames at a hopsize of
$5.8\thinspace$ms. To use Viterbi thus requires to store in memory
pointers to up to $4\thinspace$G states, which amounts to $40\thinspace$G
RAM (with uint32 python data type).

\section{Experiments}

The hopsize of computing the spectral flux feature, which resulted
in most optimal beat detection accuracy in \cite{krebs:15:ismir}
is $h_{f}=20\thinspace$ms. In comparison,\textbf{ }the hopsize of
predominant vocal melody detection is usually of smaller order i.e.
$h_{p}=5.8\thinspace$ms (corresponding to 256 frames at sampling
rate of 44100). Preliminary experiments showed that extracting pitch
with values of $h_{p}$ bigger than this values reasonably deteriorates
the vocal onset accuracy. Therefore in this work we use hopsize of
$5.8\thinspace$ms for the extraction of both features. The time difference
parameter for the spectral flux computation remains unaffected by
this change in hopsize, because it can be set separately. 

As a baseline we run the algorithm of \cite{mauch2015computer} with
the 105 note states, we introduced in Section \ref{subsec:Note-state}\footnote{We ported the original VAMP plugin implementation to python, which
is available at \linebreak{}
\href{https://github.com/georgid/pypYIN}{https://github.com/georgid/pypYIN}}. The note transition probability is the original as presented in
Eq. \ref{eq:note_trans_probs}, i.e. not aware of beats. Note that
in \cite{mauch2015computer} the authors introduce a post-processing
step, in which onsets of consecutive sung notes with same pitch are
detected considering their intensity difference. We excluded this
step in all system variants presented, because it could not be integrated
in the proposed observation model in a trivial way. This means that,
essentially, in this paper cases of consecutive same-pitch notes are
missed, which decreases inevitably recall, compared to the original
algorithm. 
\begin{table*}[t]
\begin{centering}
\subfloat{\centering{}%
\begin{tabular}{|c|c|c|c|c|c|}
\hline 
meter &  & beat Fmeas & P & R & Fmeas\tabularnewline
\hline 
\multirow{3}{*}{aksak} & Mauch & - & 33.1 & 31.6 & 31.6\tabularnewline
\cline{2-6} 
 & Ex-1 & - & 37.5 & 38.4 & 37.2\tabularnewline
\cline{2-6} 
 & Ex-2 & 86.4 & 37.8 & 36.1 & 36.1\tabularnewline
\hline 
\hline 
\multirow{3}{*}{d{\"u}yek} & Mauch & - & 42.1 & 36.9 & 37.9\tabularnewline
\cline{2-6} 
 & Ex-1 & - & 44.3 & 41.0 & 41.4\tabularnewline
\cline{2-6} 
 & Ex-2 & 72.9 & 45.0 & 39.0 & 40.3\tabularnewline
\hline 
\end{tabular}}\subfloat{\centering{}%
\begin{tabular}{|c|c|c|c|c|c|}
\hline 
meter &  & beat Fmeas & P & R & Fmeas\tabularnewline
\hline 
\multirow{3}{*}{4/4} & Mauch & - & 29.6 & 38.3 & 33.2\tabularnewline
\cline{2-6} 
 & Ex-1 & - & 30.3 & 42.5 & 35.1\tabularnewline
\cline{2-6} 
 & Ex-2 & 94.2 & 31.6 & 39.4 & 34.4\tabularnewline
\hline 
\hline 
\multirow{3}{*}{total} & Mauch & - & 34.8 & 35.6 & 35.2\tabularnewline
\cline{2-6} 
 & Ex-1 & - & 38.3 & 40.6 & 39.5\tabularnewline
\cline{2-6} 
 & Ex-2 & 84.3 & 38.1 & 38.2 & 38.1\tabularnewline
\hline 
\end{tabular}}
\par\end{centering}
\caption{Evaluation results for Experiment 1 (shown as Ex-1) and Experiment
2 (shown as Ex-2). Mauch stands for the baseline, following the approach
of \cite{mauch2015computer}. P, R and Fmeas denote the precision,
recall and f-measure of detected vocal onsets. Results are averaged
per meter type.}
\label{results_all}
\end{table*}

\subsection{Evaluation metrics}

\subsubsection{Beat detection}

Since improvement of the beat detector is outside the scope of this
study, we report accuracy of detected beats only in terms of their
f-measure\footnote{The evaluation script used is at \href{https://github.com/CPJKU/madmom/blob/master/madmom/evaluation/beats.py}{https://github.com/CPJKU/madmom/blob/master/madmom/evaluation/beats.py}}.
This serves solely the sake of comparison to existing work\footnote{Note that the f-measure is agnostic to the phase of the detected beats,
which is clearly not optimal}. The f-measure can take a maximum value of 1, while beats tapped
on the off-beat relative to annotations will be assigned an f-measure
of 0. We used the default tolerance window of $70\thinspace$ms, also
applied in \cite{holzapfel2014tracking}. %

\subsubsection{Vocal onset detection}

We measured vocal onset accuracy in terms of precision and recall\footnote{We used the evaluation script available at \href{https://github.com/craffel/mir_eval}{https://github.com/craffel/mir\_{}eval}}.
Unlike a cappella singing, the exact onset times of singing voice
accompanied by instruments, might be much more ambiguous. To accommodate
this fact, we adopted the tolerance of $t=50\thinspace$ms, used for
vocal onsets in accompanied flamenco singing by \cite{kroher2015automatic},
which is much bigger than the $t=5\thinspace$ms used by \cite{mauch2015computer}
for a cappella. Note that measuring transcription accuracy remains
outside the scope of this study.%

\subsection{Experiment 1: With manually annotated beats}

As a precursor to evaluating the full-fledged model, we conducted
an experiment with manually annotated beats. This is done to test
the general feasibility of the proposed note transition model (presented
in \ref{par:Note-transition-probability}), unbiased from errors in
the beat detection.

We did apply both the simple and the time-redistribution weighting
schemes, presented respectively in Eq. \ref{eq:weight_detected} and
in Eq. \ref{eq:weight_anno}. In preliminary experiments we saw that
with annotated beats the simple weighting yields much worse onset
accuracy than the time-redistributed one. Therefore the results reported
are conducted with the latter weighting. 

We have tested different pairs of values for $w$ and $\sigma$ from
Eq. \ref{eq:note_onset_trans_prob}. For Turkish makam the onset detection
accuracy peaks at $w=1.2$ and $\sigma=30\thinspace$ms, whereas for
the English pop optimal are $w=1.1$ and $\sigma=45\thinspace$ms.
Table 1 presents metrics compared to the baseline\footnote{Per-recording results for the makam dataset are available at \href{https://tinyurl.com/y8r73zfh}{https://tinyurl.com/y8r73zfh}
and for the \emph{lakh-vocal-segments} dataset at \href{https://tinyurl.com/y9a67p8u}{https://tinyurl.com/y9a67p8u}}. Inspection of detections revealed that the metrical-accent aware
model could successfully detect certain onsets close to beats, which
are omitted by the baseline.

\subsection{Experiment 2: Full model }

To assure computational efficient decoding, we did an efficient implementation
of the joint state space of \cite{krebs:15:ismir}\footnote{We extended the python toolbox for beat tracking https://github.com/CPJKU/madmom/,
which we make available at \href{https://github.com/georgid/madmom}{https://github.com/georgid/madmom} }. To compare to that work, we measured the beat detection with both
their original implementation and our proposed one. Expectedly, the
average f-measure of the detected beats were the same for each of
the three metrical cycle types in the datasets, which can be seen
in Table \ref{results_all}. For aksak and d{\"u}yek usuls, the accuracy
is somewhat worse than the results of $91$ and $85.2$ respectively,
reported in \cite[Table 1.a-c, R=1]{holzapfel2014tracking}. We believe
the reason is in the smaller size of our training data. Table \ref{results_all}
evidences also a reasonable improvement of the vocal onset detection
accuracy for both music traditions. The results reported are only
with the simple weighting scheme for the vocal note onset transition
model (the time-redistribution weighting was not implemented in this
experiment). 

Adding the automatic beat tracking improved the baseline, whereas
this was not the case with manual beats for simple weighting. This
suggests that the concurrent tracking of beats and vocal onsets is
a flexible strategy and can accommodate some vocal onsets, slightly
time-shifted from a beat. We observe also that the vocal onset accuracy
is on average a bit inferior to that with manual beat annotations
(done with the time-redistribution weighting). 

For the 4/4 meter, despite the highest beat detection accuracy, the
improvement of onset accuracy over the baseline is the least. One
reason for that may be that the note probability pattern $e(b)$,
used for 4/4 is not well representative for the singing style differences.

A paired t-test between the baseline and each of Ex-1 and Ex-2 resulted
in p-values of respectively $0.28$ and $0.31$ on total for all meter
types. We expect that statistical significance can be evaluated more
accurately with a bigger number of recordings. 

\section{Conclusions}

\textcolor{black}{In this paper we presented a Bayesian approach for
tracking vocal onsets of singing voice in polyphonic music recordings.
The main contribution is that we integrate in one coherent model two
existing probabilistic approaches for different tasks: beat tracking
and note transcription.}\textcolor{red}{{} }Results confirm that the
knowledge of the current position in the metrical cycle can improve
the accuracy of vocal note onset detection over different metrical
cycle types. The model has a comprehensive set of parameters, whose
appropriate tuning allows application to material with different singing
style and meter. 

In the future the manual adjustment of these parameters could be replaced
by learning their values from sufficiently big training data, which
was not present for this study. In particular, the \emph{lakh-vocal-segments
}dataset could be easily extended substantially, which we plan to
do in the future. Moreover, one could decrease the expected parameter
values range, based on learnt values, and thus decrease the size of
the state space, which is a current computational limitation. \textcolor{black}{We
believe that the proposed model could be applied as well to full-fledged
transcription of singing voice.}

\subparagraph*{Acknowledgements}

We thank Sebastian B{\"o}ck for the implementation hints. Ajay Srinivasamurthy
is currently with the Idiap Research Institute, Martigny, Switzerland. 

This work is partly supported by the European Research Council under
the European Union\textquoteright s Seventh Framework Program, as
part of the CompMusic project (ERC grant agreement 267583) and partly
by the Spanish Ministry of Economy and Competitiveness, through the
\textquotedblright Mar{\'i}a de Maeztu\textquotedblright{} Programme
for Centres/Units of Excellence in R\&D\textquotedblright{} (MDM-2015-0502).

\bibliographystyle{plain}
\bibliography{JabRefOnsetDetectionFullReferences}

\end{document}